\documentclass[aip,twocolumn,letterpaper]{revtex4}

\usepackage{fullpage}

\usepackage{graphics}
\usepackage{epsfig}

\usepackage[margin=1.0in]{geometry}
\usepackage{xcolor}
\definecolor{ultramarine}{rgb}{0.07, 0.04, 0.56}

\begin{document} 

\title{Electric field effects during disruptions }
\author{Allen H Boozer}
\affiliation{Columbia University, New York, NY  10027 \linebreak ahb17@columbia.edu}

\begin{abstract} 
Tokamak disruptions are associated with breaking magnetic surfaces, which makes magnetic field lines chaotic in large regions of the plasma.  The enforcement of quasi-neutrality in a region of chaotic field lines requires an electric potential that has both short and long correlation distances across the magnetic field lines.  The short correlation distances produce a Bohm-like diffusion coefficient $\sim T_e/eB$ and the long correlation distances $a_T$ produce a large scale flow $\sim  T_e/eB a_T$.  This cross-field diffusion and flow are important for sweeping impurities into the core of a disrupting tokamak.  The analysis separates of the electric field in a plasma into the sum of a divergence-free, $\vec{E}_B$, and a curl-free, $\vec{E}_q$, part, a Helmholtz decomposition.  The divergence-free part of $\vec{E}$ determines the evolution of the magnetic field.  The curl-free part  enforces quasi-neutrality, $\vec{E}_q=-\vec{\nabla}\Phi_q$.  Magnetic helicity evolution gives the required boundary condition for a unique Helmholtz decomposition and an unfortunate constraint on steady-state tokamak maintenance.  \color{black}

\end{abstract}

\date{\today} 
\maketitle

\section{Introduction} 

A tokamak disruption generally involves a large-scale breaking of the magnetic surfaces on a very short time scale, approximately a thousand times shorter than the timescale for the resistive decay of the plasma current.  Such disruptions can be identified with a fast magnetic reconnection, which is discussed in Section \ref{B-rec}.  

As the breaking of magnetic surfaces proceeds, the field lines transition from lying on surfaces to being chaotic.  A chaotic magnetic field line has two important properties:  (i)  A chaotic line has lines infinitesimally separated from it that have a separation that increases exponentially $e^{\sigma(\ell)}$ with distance $\ell$ along the line.  (ii) A chaotic line comes arbitrarily close to every point in chaotic region as $\ell\rightarrow\infty$.  A single line is said to cover an entire region of chaotic field lines, but the number of toroidal transits the line must make to even crudely sample the entire region is generally extremely large.

Once a region of magnetic surfaces is transformed into a chaotic region, the magnetic field and the plasma pressure quickly reestablish force balance across the magnetic field lines.  Force balance is reestablished on a much longer timescale along the lines due to the difference in length scales.  The length scale across the lines is the minor radius $a$.  The length scale along the lines is the many toroidal transits, each of length $2\pi R$, that are required for a single line to adequately cover a chaotic region for the smoothing effects of cross-field transport to be important.

Electrons move far faster along magnetic field lines than ions. As long as a pressure variations persist along magnetic field lines, an electric potential $\Phi_q(\ell)$ is required along each line to preserve quasi-neutrality.  The magnetic field lines that are close to each other at one point in chaotic region separate as $e^{\sigma(\ell)}$.  The length scale over which $\Phi_q(\ell)$ changes is determined by the scale along the magnetic field lines the pressure changes, which is generally very long compared to the length scale over which the exponentiation $\sigma(\ell)$ increases.  

As discussed in Section \ref{sec:Plasma-ev}, the difference in scale in $\ell$ of $\Phi_q$ and the exponentiation $\sigma$ implies that quasi-neutrality produces a large $\vec{E}\times\vec{B}$ drift, $\vec{B}\times\vec{\nabla}\Phi_q$ across the magnetic field with both short and long-scale correlation distances.  The short scale correlation gives a diffusion that is Bohm like $D_q \approx \big<T_e\big>/eB$.  The electron temperature on the scale that $\Phi_q(\ell)$ is field line dependent determines $\big<T_e\big>$.  There is also a large scale flow $\approx T_e/(eBa_T)$, where $a_T$ is the radial distance scale over which electron temperature changes, which is generally comparable to the minor radius $a$.  

The diffusion and flow produced by the quasi-neutrality electric potential spreads impurities across the plasma.  High-Z ions move slowly along magnetic field lines but have the same $\vec{E}\times\vec{B}$ drifts as other species.

 The importance of magnetic field line chaos to transport has been broadly appreciated since the 1978 paper of Rechester and Rosenbluth \cite{Rechester-Rosenbluth}.  Their paper was on electron heat transport along chaotic field lines and not on the transport across the magnetic field lines produced by the electric field required for quasi-neutrality.  This distinction is essential for understanding the effect of a chaotic magnetic field on impurity transport during disruptions.  Impurities move at a far slower speed than electrons along magnetic field lines, but all species have the same $\vec{E}\times\vec{B}/B^2$ velocity across the magnetic field.

Section \ref{sec:Helmholtz} discusses the two remarkably independent parts of the electric field in a plasma.  One part gives the evolution of the magnetic field, and the other the $\vec{E}\times\vec{B}$ drift of the plasma across the magnetic field lines.  These two parts are uniquely separated by a Helmholtz decomposition of the electric field, which separates a vector in three-space into the sum of a divergence-free and a curl-free part.  A Helmholtz decomposition is unique if a boundary condition is given on an enclosing surface.  In a toroidal plasma, the chamber wall provides the boundary condition.  

The determination of the correct boundary condition on the wall is subtle, but can be resolved with the use of the equation for the evolution of the magnetic helicity enclosed by the wall, Appendix \ref{sec:helicity}.   Toroidal plasma equilibria require a non-zero helicity content.  In a tokamak this is produced by the toroidal current, which resistively decays.  By controlling the electric potential on the chamber wall, the helicity content could be held constant.  Unfortunately, the resulting tokamak would have magnetic surfaces that are periodically destroyed as in a major disruption, Appendix  \ref{sec:helicity}.

Section \ref{sec: Simulations} discusses simulations of disruptions, and Section \ref{sec: Discussion} is a summary.


\section{Magnetic Field Evolution}

\subsection{General expression for a magnetic evolution}

The evolution of magnetic fields is given by Faraday's law, $\partial\vec{B}/\partial t= -\vec{\nabla}\times \vec{E}$, which can be placed in a clarifying form by the general representation in three dimensions of a vector field $\vec{E}$ in terms of another vector field $\vec{B}$,
\begin{equation}
\vec{E} = - \vec{u}_\bot\times\vec{B} - \vec{\nabla}\Phi + \frac{V_\ell}{2\pi}\vec{\nabla}\varphi.  \label{E eq}
\end{equation} 

The expressions that appear in this equation have physical interpretations.  $V_\ell$ is the loop voltage in toroidal plasmas with $\varphi$ the toroidal angle.  The loop voltage is constant along a field line but is required to ensure the potential $\Phi$ is single valued---a necessary condition for $\vec{\nabla}\times(\vec{\nabla}\Phi)$ to give a zero loop integral using Stokes' theorem.  

The velocity $\vec{u}_\bot$ is the velocity of the magnetic field lines when the loop voltage is zero, as demonstrated by Newcomb \cite{Newcomb}, and defines a convenient magnetic frame of reference when it is not.  The concept of $\vec{u}_\bot$ seems essential to understand what can and what cannot produce a breaking of magnetic field lines \cite{Boozer:RMP}.   Nevertheless, both Newcomb \cite{Newcomb} and Greene \cite{Greene:1993} have stated that the concept of a magnetic field line velocity has no physical content, and the concept is little known or understood by plasma physicists.

\color{black}  

 Equation (\ref{E eq}) is important, not well known, but easily proven.  The component of $\vec{E}$ along $\vec{B}$ is $E_{||} = -\partial \Phi/\partial \ell + (V_\ell/2\pi) \partial\varphi/\partial \ell$ with $\ell$ the distance along a field line;  $\vec{B}\cdot\vec{\nabla}\Phi = B\partial\Phi/\partial \ell$.  The two components of $\vec{E}$ perpendicular to $\vec{B}$ are represented using $\vec{u}_\bot$.  Magnetic reconnection occurs if and only if  the gradient of the \color{black} required  loop voltage  $V_\ell$ for a single-valued potential $\Phi$ is non-zero.  The best known contribution to $V_\ell$ is the  toroidal loop integral of $\eta_{||} j_{||}$, which is $V_\ell=\eta_{||}j_{||} 2\pi R_0$ at the magnetic axis.

The general magnetic evolution, which is given by Farday's Law, can be written as
\begin{eqnarray}
\frac{\partial\vec{B}}{\partial t} = -\vec{\nabla}\times(\vec{u}_\bot \times\vec{B}) - \vec{\nabla}V_{\ell}\times\vec{\nabla}\frac{\varphi}{2\pi}. \label{B-ev}
\end{eqnarray}

The importance of this equation is illustrated by the demonstration that the Hall term in Ohm's law is of no direct relevance to reconnection.  The Hall term is $\vec{f}_L/en$, where the Lorentz force $\vec{f}_L =\vec{j}\times\vec{B}$, and appears on the right-hand side of a generalized Ohm's Law, Equation (\ref{Hall-Ohm's}).   The Hall term is often considered to be essential for fast magnetic reconnection, as in the often cited 2001 review of reconnection in two-dimensional systems by Birn et al \cite{reconnection review}, but it does not enter Equation (\ref{B-ev}) for the evolution of magnetic fields and is certainly not required for fast magnetic reconnection when the magnetic field depends non-trivially on all three spatial coordinates.  

The resolution of the paradox of fast reconnection is simple in three-dimensional systems, magnetic field line chaos \cite{Boozer:B-ev}, which is discussed in Section \ref{B-rec}.   The reason magnetic chaos requires three dimensions is that when the vector potential is written in the form of Equation (\ref{A}), the field lines are given by the poloidal flux $\psi_p(\psi,\theta,\varphi)$.  The poloidal flux is the Hamiltonian \cite{Boozer:RMP} for the field lines $d\psi/d\varphi =-\partial \psi_p/\partial \theta$, $d\theta/d\varphi =\partial \psi_p/\partial \psi$ and $d\psi_p/d\varphi=\partial\psi_p/\partial\varphi$. When $\psi_p$ is independent of any of the three variables, the conjugate variable is a constant of the motion.  For example, when $\psi_p$ is independent of $\varphi$ a field line remains on the same $\psi_p$ surface forever---no matter how complicated that surface may be.   Even when $\psi_p$ is a simple analytic function that depends on all three coordinates it can, and generally does, give neighboring field lines that exponentiate apart throughout regions of space.  A simple Hamiltonian \cite{Ham-sep} defines the general conditions for a arbitrary magnetic field line to have lines infinitesimally separated from it which exponentially increase their separation with distance along the line.


\subsection{Magnetic reconnection \label{B-rec}} \color{black}

Magnetic surfaces break on an arbitrarily short timescale compared to the global resistive timescale when the separation between neighboring magnetic surfaces varies by an exponentially large factor \cite{Boozer-surf-dest:2022}.  The loop voltage need only cause field-line diffusion across the shortest distance between magnetic surfaces to cause a large scale topological change from nested surfaces to chaotic field lines.  A large current density is not required.

\begin{figure}
\centerline { \includegraphics[width=3.0in]{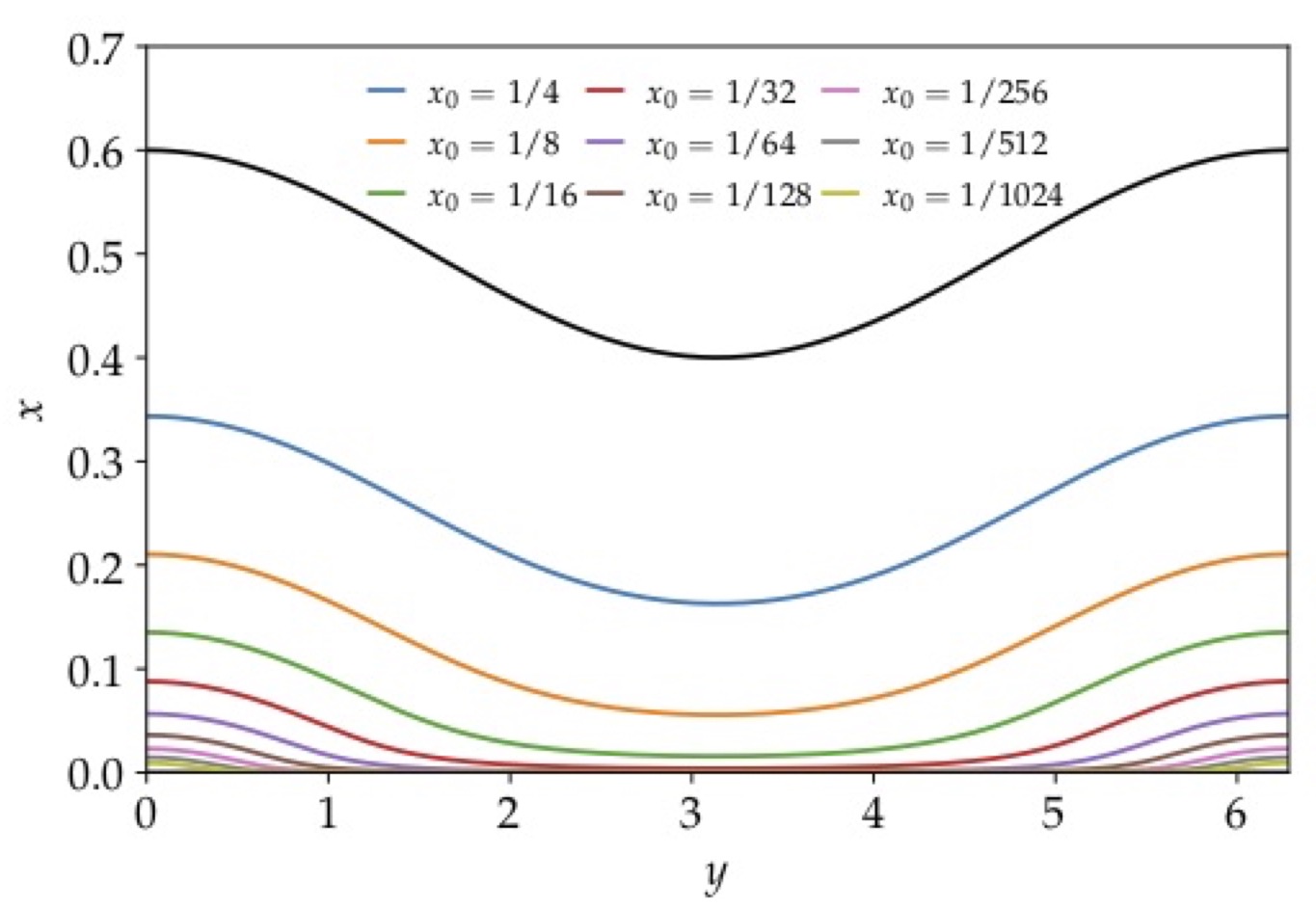} }
\caption{This figure gives the perturbed shape of magnetic surfaces that had different initial separations $x_0$ from the rational surface $\iota(\psi_s)=M_s/N_s$ in response to a perturbation $\cos(M_s\theta-N_s\varphi)$ imposed far from that surface.  The distance of perturbed  surfaces from the rational surface, $x$, is normalized by the amplitude of the perturbation just outside the current channel near the rational surface, and  $y\equiv M_s\theta-N_s\varphi$.  Reproduced from Y.-M. Huang, S. R. Hudson, J. Loizu, Y. Zhou, and A. Bhattacharjee,  Phys. Plasmas \textbf{29}, 032513 (2022).  }  
\label{fig:Displacement}
\end{figure}

As shown by Huang, Hudson,  Loizu, Zhou, and Bhattacharjee \cite{Res-ideal2022}, Figure \ref{fig:Displacement}, an ideal magnetic perturbation presses magnetic surfaces close together near a resonant rational surface at the location where an X-point, or more correctly an X-line, would arise in a curl-free magnetic field.  Since the magnetic flux between magnetic surfaces is conserved in an ideal evolution, the separation between neighboring surfaces must be large elsewhere.   Where two neighboring magnetic field lines are close to each other on a magnetic surface, which means $\Delta\theta$ small, the magnetic surfaces are far apart, which means the $\Delta\psi$ between the surfaces  is large, and vice-versa \cite{Boozer-surf-dest:2022}.   When an ideal perturbation makes magnetic surfaces highly contorted, the neighboring field lines in those surfaces can exponentiate apart an arbitrarily large amount.  This exponentiation goes up and down so the Lyapunov exponent, the average rate of exponentiation, is zero.  In 2022, Jardin et al \cite{Jardin:fast breakup:2022} reported the breakup of magnetic surfaces in M3D-C1 simulations on a timescale set by an ideal MHD instability.

Equation (\ref{B-ev}), which contains $\vec{u}_\bot$ and the loop voltage $V_\ell$, is the general evolution equation for the magnetic field.  But, a related equation that uses the simple Ohm's law, $\vec{E}+\vec{v}\times\vec{B}=\eta\vec{j}$, with $\eta$ constant, plus Ampere's law makes it more explicit that the magnetic evolution equation is of the advection-diffusion form

\begin{equation}
\frac{\partial\vec{B}}{\partial t}=\vec{\nabla}\times(\vec{v}\times\vec{B}) + \frac{\eta}{\mu_0}\nabla^2\vec{B}.
\end{equation}
It has been known since the 1984 paper of Aref \cite{Aref:1984} that when the flow $\vec{v}$ is chaotic that the diffusion, $\eta/\mu_0$, causes dissipative relaxation on the ideal timescale multiplied by a factor that depends only logarithmically on the diffusive divided by the ideal timescale.  The ideal timescale is $\tau_i\equiv L/|\vec{v}|$, where $L$ is the spatial scale of the system and  $|\vec{v}|$ is the flow speed of the ideal evolution.  The diffusive time scale is $\tau_d=(\mu_0/\eta)L^2$.  The magnetic Reynolds number is defined as $R_m\equiv\tau_d/\tau_i$; typically $\ln R_m$ is of order ten in large tokamaks

Where magnetic surfaces are pressed  together,  $\nabla^2\vec{B}$ increases until a local breaking of the flux surfaces occurs.  Although $\eta$ must be non-zero where the surfaces are pressed together to have reconnection, at other locations the magnetic field evolves essentially ideally, as if $\eta$ were zero.  

When the breakup of magnetic surfaces occurs on a short timescale compared to the global resistive-evolution timescale, the magnetic evolution is accurately given by
\begin{equation} 
\vec{E} = - \vec{u}_\bot\times\vec{B}-\vec{\nabla}\Phi \label{Ideal B ev}
\end{equation} 
almost everywhere.  In three dimensional, but not in two dimensional cases, magnetic field lines reconnect in highly localized places along the lines, but this can change the field line topology everywhere.  As explained, magnetic field line chaos is not mathematically possible in two-dimensional systems.

\begin{figure}
\centerline { \includegraphics[width=3.0in]{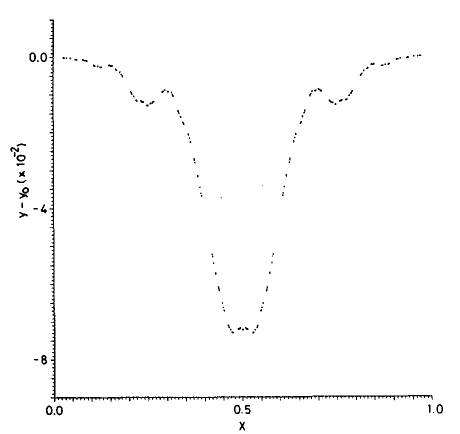} }
\caption{The irrational surface that has a transform that is the most difficult to approximate by a rational number is the last surface to break.  That surface becomes a Cantorus, which resembles an irrational surface but has gaps, called turnstiles, through which magnetic flux passes. This illustration of a Cantorus is from R. S. MacKay, J. D. Meiss, and I. C. Percival, Physica \textbf{13D}, 55 (1984).  }  
\label{fig:Cantorus}
\end{figure}

\color{black}

Important studies of the simulations of magnetic surface breakup remain to be done, Section \ref{sec: Simulations}.  What can be said is that  the breakup begins at low order rational surfaces but does not produce a large chaotic field-line region until the last irrational magnetic surface between the low order rational surfaces is broken.  In 1984, MacKay, Meiss, and Percival  \cite{Cantorus:1984} studied the breakup of the last irrational surface using the standard map.  They found the last irrational surface breaks to form a Cantorus, which has small holes or slits, called turnstiles, through which two tubes of equal magnetic flux---one inward and one outward---can cross.    As they noted: ``Most of the gaps in a Cantorus are very small, since their total length is finite. Even when there are large gaps, orbits can take a long time to get through."  A field line followed on a Cantorus can make a large number of toroidal circuits and appear to lie on a magnetic surface, but eventually the line will leave the Cantorus and enter a chaotic sea of field lines through a turnstile.

\color{black}

The timescale for an evolving magnetic field to change its topology from having confining magnetic surfaces to large scale chaos is a critical issue for determining the damage from runaway electrons to tokamaks.  In 2016, Boozer and Punjabi \cite{Boozer-Punjabi} found that runaway electrons filling a chaotic core in a post-disruption tokamak could strike the chamber walls in either an extremely short or very long spatial and temporal scale.  The determining factor was whether the confining annulus of magnetic surfaces was broken slowly or rapidly compared to the time required for a relativistic electron to cover the central chaotic field-line region.  The spatial-temporal concentration produced by a slow breaking would be disastrous in a power-plant scale tokamak.  The spatial-temporal spreading given by a rapid breaking would result in negligible damage.   The effects of a concentrated loss have been observed in many tokamak experiments.  In 2021 two experiments \cite{Reux:2021,Soldan:2021} showed the benign effect of runaway losses when the confining annulus is rapidly broken.

The evolution equations for thermodynamic quantities, such as the mass density $\rho$ and the entropy per unit volume $s$, have the mass flow velocity $\vec{v}$ in the advective term of their evolution equations.  The evolution equation for the mass density is archetypal,
\begin{equation}
\frac{\partial\rho}{\partial t}+\vec{\nabla}\cdot(\rho\vec{v})=\vec{\nabla}\cdot(D\vec{\nabla}\rho). \label{density-ev}
\end{equation}
The diffusive flux $-D\vec{\nabla}\rho$ can be far more complicated that this simple form and represents density transport effects on the gryro-radius scale, such as neoclassical or micro-turbulent transport, or small scale turbulent eddies in the mass flow that are not explicitly resolved.  As always \cite{Aref:1984} a chaotic velocity in an advection-diffusion equation makes the ideal, $D=0$, solution relevant only for a timescale that depends logarithmically on the diffusion coefficient as $D\rightarrow0$.  

A chaotic flow $\vec{v}$ in the thermodynamic evolution equations does not necessarily imply a chaotic flow $\vec{u}_\bot$ in Equation (\ref{B-ev}) for the magnetic field evolution for two reasons:  (1) The plasma flow includes a flow along the magnetic field lines $\vec{v}_{||}$ but the magnetic field line flow $\vec{u}_\bot$ does not.  (2) The flows perpendicular to the magnetic field differ with $\vec{v}_\bot-\vec{u}_\bot$ given by Equation (\ref{Plasma-B vel}).


\section{The Helmholtz decomposition of the electric field \label{sec:Helmholtz} } 

In plasmas, the electric field has two remarkably-independent parts.  One gives the evolution of the magnetic field, and the other the $\vec{E}\times\vec{B}$ drift of the plasma across the magnetic field lines.  

These two parts are uniquely separated by a Helmholtz decomposition of the electric field, which separates a vector in three-space into the sum of a divergence-free and a curl-free part.  A Helmholtz decomposition is unique if a boundary condition is given on an enclosing surface.  In a toroidal plasma, the chamber wall provides the boundary condition.  

The determination of the correct boundary condition on the wall is subtle, but can be resolved with the use of the equation for the evolution of the magnetic helicity enclosed by the wall, Appendix \ref{sec:helicity}.   Toroidal plasma equilibria require a non-zero helicity content.  In a tokamak this is produced by the toroidal current, which resistively decays.  By controlling the electric potential on the chamber wall, the helicity content could be held constant.  Unfortunately, the resulting tokamak would have magnetic surfaces that are periodically destroyed as in a major disruption, Appendix  \ref{sec:helicity}.

The interpretation of each of the two parts of $\vec{E}$ is given by two of Maxwell's equations, $\vec{\nabla}\cdot\vec{E} = \rho_{ch}/\epsilon_0$ and $\vec{\nabla}\times \vec{E} = -\partial\vec{B}/\partial t$.  The curl-free part of the electric field, which will be denoted by $\vec{E}_q$ has a divergence that equals the net charge density $\rho_{ch}$.  In a hydrogenic plasma, $\vec{\nabla}\cdot\vec{E}_q =e(n_h-n_e)/\epsilon_0$.  Inexact charge neutrality causes $\vec{E}_q$ to become as large as necessary to preserve quasi-neutrality when the Debye length is short compared the spatial scales in the plasma.   Since $\vec{\nabla}\times \vec{E}_q=0$, this part of the electric field can be written as the gradient of a well behaved scalar potential, 
\begin{eqnarray}
\vec{E}_q&=&-\vec{\nabla}\Phi_q;\\
\nabla^2\Phi_q&=&-\frac{e(n_h-n_e)}{\epsilon_0};\\
\Phi_q]_{wall}&=&0.
\end{eqnarray}
Appendix \ref{sec:helicity} shows $\Phi_q]_{wall}=0$ is the correct boundary condition.  This boundary condition requires that  $\Phi_q$ include the Debye-length scale sheath potential that arises near the wall.  The sheath potential is changed little by the current density $j_{||}$ flowing along the magnetic field when $|j_{||}|<<j_s$.  The ion saturation current, $j_s\equiv enC_s\approx 0.1 n_{20} \sqrt{T_e}$~MA/m$^2$, where the electron density is in units of $10^{20}/$m$^3$ and the temperature is in electron-Volts.

The divergence-free part of the electric field, which will be denoted by 
\begin{equation}
\vec{E}_B\equiv \vec{E}+\vec{\nabla}\Phi_q,
\end{equation} 
is the only part that is relevant to the evolution of the magnetic field: the movement of the magnetic field lines and their breaking.  In the Coulomb gauge, $\vec{\nabla}\cdot\vec{A}=0$, the magnetic evolution is given by $\vec{E}_B=-\partial \vec{A}/\partial t -\vec{\nabla}\Phi_B$.  Since $\vec{\nabla}\cdot\vec{E}_B=0$, the potential $\Phi_B$ must satisfy $\nabla^2\Phi_B=0$.  The plasma charge density, which gives $\Phi_q$, has no affect on $\Phi_B$.

Not only is the electric field the sum of two parts, $\vec{E}=\vec{E}_B+\vec{E}_q$, but the plasma velocity is as well.  It is the sum of the velocity of the magnetic field lines $\vec{u}_\bot$, as discussed in connection with Equation (\ref{E eq}), and the velocity of the plasma relative to the field lines.   

\color{black}

The velocity $\vec{v}$ of the plasma is defined by its mass flow and is related to other physical quantities by Ohm's law.  Including the Hall term, Ohm's law is
\begin{equation}
 \vec{E} + \vec{v}\times\vec{B}= \eta_{||}\vec{j}_{||}+\eta_\bot\vec{j}_{\bot} +\frac{\vec{f}_L}{n e}, \label{Hall-Ohm's}
\end{equation}
where $\vec{f}_L=\vec{j}\times\vec{B}$ is the Lorentz force and $n$ is the electron number density.  The velocity of the plasma consists of a part $\vec{v}_{||}$  along the magnetic field lines and a part across.  The difference between $\vec{v}_\bot$ and $\vec{u}_\bot$ is obtained by substituting the electric field from Equation (\ref{E eq}) into Equation (\ref{Hall-Ohm's}) to obtain
\begin{eqnarray}
(\vec{v}_\bot-\vec{u}_\bot)\times\vec{B}= && \vec{\nabla}_\bot\Phi +\frac{\vec{f}_L}{n_e e} \nonumber\\
 && +\eta_\bot\vec{j}_{\bot} -\frac{V_\ell}{2\pi}\vec{\nabla}_\bot\varphi. \hspace{0.2in} \label{Plasma-B vel}
\end{eqnarray}
The difference between the plasma flow across the magnetic field and the field lines themselves is due both to the ideal terms, which are in the upper line of the right-hand side of Equation (\ref{Plasma-B vel}), and the dissipative terms, which are given in the lower line.

  During an ideal evolution of the magnetic field, $\vec{u}_\bot $ is the velocity of the magnetic field lines, and $\vec{v}-\vec{u}_\bot$ is the velocity of the plasma relative to the magnetic field, whether the plasma is responding ideally or not.  In particular, the quasi-neutrality electric field, $\vec{E}_q=-\vec{\nabla}\Phi_q$, gives an $\vec{E} \times\vec{B}$ drift across the magnetic field
\begin{equation}
\vec{V}_q = \frac{\vec{B}\times\vec{\nabla}\Phi_q}{B^2}.
\end{equation}


\section{Diffusion and flows associated with quasi-neutrality \label{sec:Plasma-ev} }

In plasmas of fusion interest, the smallness of the Debye length compared to spatial scales gives the quasi-neutrality constraint, the fractional difference between the number of positive and negative charges must be far smaller than unity.  The quasi-neutrality constraint is enforced by the potential $\Phi_q$, which gives both an $\vec{E}\times\vec{B}$ flow on the large scale variation of the electron temperature across the magnetic lines, $a_T$, and in a chaotic field line region a Bohm-like diffusion due to the  short correlation scale  of the electron temperature $T_e$ across the magnetic field in a region of chaotic magnetic field lines \color{black}  coupled with the rapid equilibration of $T_e$ along the magnetic field lines.  Both effects are considered in this section.  

The ideal response of hydrogenic ions and electrons is related to the electric field by the equations
\begin{eqnarray} 
\vec{E} &=& - \vec{v}_h\times\vec{B} + \frac{1}{e}\left(m_h \frac{d\vec{v}_h}{dt} -\frac{\vec{\nabla}\cdot\tensor{p}_h}{n}\right); \label{ion eq}\\
&=& - \vec{v}_e\times\vec{B} +\frac{1}{e}\frac{\vec{\nabla}p_e}{n}, \label{electron eq}
\end{eqnarray}  
where the small electron-inertia term has been neglected for simplicity.   The ion pressure tensor is used in order to include the effect of viscosity.    The quasi-neutrality constraint implies the number density of the ions and electrons are equal to $n$ to high accuracy.

The curl of Equation (\ref{electron eq}) is interesting.  When $(\vec{v}_e\times\vec{B}) =(\vec{u}_\bot\times\vec{B}) + \vec{\nabla}\Phi_q$, one finds that \begin{equation} 
\vec{\nabla}\times(\vec{E}+\vec{u}_\bot\times\vec{B})=-\frac{\vec{\nabla}n\times\vec{\nabla}T_e}{en^2}.
\end{equation}
A non-zero $\vec{\nabla}n\times\vec{\nabla}T_e$ is called a Biermann battery \cite{Biermann}.  Although the Biermann battery term is non-zero in a region of chaotic field lines, it is generally assumed to be negligibly small.

  Equation (\ref{ion eq}) for the ions can be replaced by the difference between Equations (\ref{ion eq}) and (\ref{electron eq}), which gives the standard force balance equation,  
\begin{equation}
m_h \frac{d\vec{v}_h}{dt} = \vec{j}\times\vec{B} - \vec{\nabla}\cdot \tensor{p}.
\end{equation}
The current density is $j=en(\vec{v}_h-\vec{v}_e)$, and the pressure tensor is $\tensor{p}=\tensor{p}_h+p_e\tensor{1}$.   

The component of the electric field parallel to the magnetic field from Equations (\ref{Ideal B ev}) and (\ref{electron eq}) gives a differential equation for the potential   required for quasi-neutrality, $\Phi_q$:
\begin{equation}
\frac{\partial \Phi_q}{\partial\ell} = - \frac{1}{en}\frac{\partial p_e}{\partial\ell} \label{Phi-der}
\end{equation}
When the plasma pressure $p$ is a scalar, $p_e= pT_e/(T_h+T_e)$, so even a single-fluid MHD code can be used to obtain the potential required for quasi-neutrality.   

When the units of temperature are chosen so $p_e=nT_e$. 
\begin{eqnarray}
\frac{\partial \Phi_q}{\partial\ell} = - \frac{1}{e}\left( \frac{\partial \ln n}{\partial \ell} T_e + \frac{\partial T_e}{\partial\ell}\right), 
\end{eqnarray}
When $|\partial T_e/\partial \ell| >> T_e | \partial \ln{n}/\partial \ell |$, the potential $\Phi_q = (c_0 - T_e(\ell))/e$ with $c_0$ a constant with respect to $\ell$ but differs from one field line to another.   When the inequality is the other way around, $\Phi_q = - \ln(n(\ell)/n)_0) T_e/e$ with $n_0$ a constant with respect to $\ell$ that differs from one field line to another.  

For all expressions for $p_e(\ell)$, the value of the potential at a location within a chaotic field line region can be written as
\begin{equation}
\Phi_q = - \frac{\big< T_e \big>}{e},
\end{equation}
 where $\big< T_e \big>$ is determined by integrating Equation (\ref{Phi-der}) and changes on a short scale across the magnetic field because of the exponential sensitivity of the field line trajectories.
 
 \color{black}

The magnitude of $\Phi_q$ rises to whatever magnitude is necessary to enforce quasi-neutrality.  That magnitude is approximately $T_e/e$, so $\nabla^2 \Phi_q\approx T_e/ea_g^2$ with $a_g$ the gradient scale length.  The breaking of charge neutrality is small but non-zero; $(n_h-n_e)/n_e \approx (\lambda_d/a_g)^2$ where $\lambda_d \equiv \sqrt{(\epsilon_0 T_e/e^2n)}$ is the Debye length. 

 The $T_e/e$ variation in the potential is familiar to plasma physics from the adiabatic electron response.  The name adiabatic comes from the fact that collisions do not relax an electron distribution function $f_e\propto\exp\big((\frac{1}{2}m_ev_e^2 -e\Phi)/T_e\big)$.  No entropy is produced regardless of the variation of $\Phi$, but variations in $\Phi$ do produce variations in the electron density, which are comparable to the density itself when $\delta\Phi\sim T_e/e$.

A single magnetic field line in a bounded region of chaotic magnetic field lines comes arbitrarily close to every point in that region.  In principle, an arbitrary starting point can be used to find $\Phi_q$ throughout the region.  The value of potential and the temperature at the starting point just adds a constant to $\Phi_q$ and is irrelevant to the calculation of the diffusive transport and the flow that is caused by $\vec{\nabla}\Phi_q$ in a bounded region of chaotic magnetic field lines.  

When the chaotic field line region extends to the chamber walls, the calculation requirers the boundary condition on the walls, $\Phi_q]_w=0$, must be enforced.  Qualitatively the diffusion and flow remain the same, but showing this is subtle.  When a field line enters the plasma from the wall, the line must by magnetic flux conservation restrike the wall and will generally do so at a different value of the potential.  Let $\Phi(\ell)$ be the solution for the potential along a field line integrated from $\Phi(0)=0$, the point of entry.  The value of the potential when the line restrikes the wall after going a distance $L$  is $\Phi(L)$.  This is a loop voltage that drives a current and is part of the potential $\Phi_B$ associated with the magnetic evolution.  This part of the potential satisfies $\nabla^2\Phi_B=0$ in the Coulomb gauge.   $\Phi(L)$ at even nearby strike points will, because of the chaos, be produced along a field line with a very different trajectory and have a different value.  Each term in a Fourier decomposition of  $\Phi(L)$ at strike points will give a contribution  to the Laplacian $\nabla^2_F \Phi(L)=-k^2_s \Phi(L)$ with the important Fourier terms having large values of $k_s$.  The total Laplacian must be zero, so each Fourier term of $\Phi(L)$ must decay as $e^{-k_s\Delta\ell}$, where $\Delta\ell$ is the distance into the plasma.  The implication is that the $\Phi_B$ part of $\Phi$ is non-zero only in a very narrow boundary layer or sheath.  The two ends of the line presumably are symmetric, so away from the wall the quasi-neutrality potential can be taken to be $\Phi_q(\ell) = \Phi(\ell) - \Phi(L)/2$.   

\color{black}

The electric potential required for quasi-neutrality, $\Phi_q$, gives an $\vec{E}\times\vec{B}$ drift, $\vec{V}_q \equiv (\vec{B}\times\vec{\nabla}\Phi_q)/B^2$ and an effective diffusion coefficient, $D_q \approx |\vec{V}_q|^2\tau_{corr}$.  The correlation time of the $\vec{E}\times\vec{B}$ drift is $\tau_{corr} \approx \Delta/ |\vec{V}_q|$ with $\Delta$ the spatial scale across the field lines over which the potential is correlated.  Consequently, $D_q\approx |\vec{V}_q|\Delta\approx |\Delta\Phi_q|/B$ with $|\Delta\Phi_q|$ the expected variation in $\Phi_q$ across the magnetic field lines, which is approximately  $\big<T_e\big>/e$.  

The approximation that $\Delta\big<T_e\big> \approx \big<T_e\big>$ across the magnetic field comes from $\big<T_e\big>$ having a slow variation along each field line, but a variation remains as long as pressure differences remain across the chaotic regions in a plasma.  $\big<T_e\big>$ has a rapid variation across the magnetic field lines because chaos implies neighboring magnetic field lines have large differences in their trajectories.  Each trajectory samples the full chaotic region but in a very different way.  

The expected transport produced by the electric potential required for quasi-neutrality is then approximately a Bohm-like diffusion, $D_{B\ell}$;
\begin{eqnarray}
D_q &\approx& D_{B\ell} \label{diffusion} \\
D_{B\ell}&\equiv& \frac{\big<T_e\big>}{eB}  \\
&=& 10^3 \frac{T_{\mbox{kev}}}{B_{\mbox{Tesla}}}~\frac{\mbox{m}^2}{\mbox{s}}.
\end{eqnarray} \color{black}

A large scale flow is also produced on the scale of the distance $a_T$ between generally hot and generally cold regions---presumably the central and the edge regions.  The large scale flow has a characteristic speed
\begin{eqnarray}
V_q &\approx& \frac{\Phi_q}{a_T B}  \\
&\approx& \frac{D_{B\ell}}{a_T}. \label{flow}
\end{eqnarray}

The diffusion coefficient given in Equation (\ref{diffusion}) and the flow given in Equation (\ref{flow}) are clearly only estimates.  Great precision should not be expected.  However, their closeness to simulation results, Section \ref{sec: Simulations}, implies careful studies of $\Phi_q$ from field line integrations to determine more reliable spatial scales would be worthwhile.



\section{Simulations \label{sec: Simulations}} 

\begin{figure}
\centerline { \includegraphics[width=3.0in]{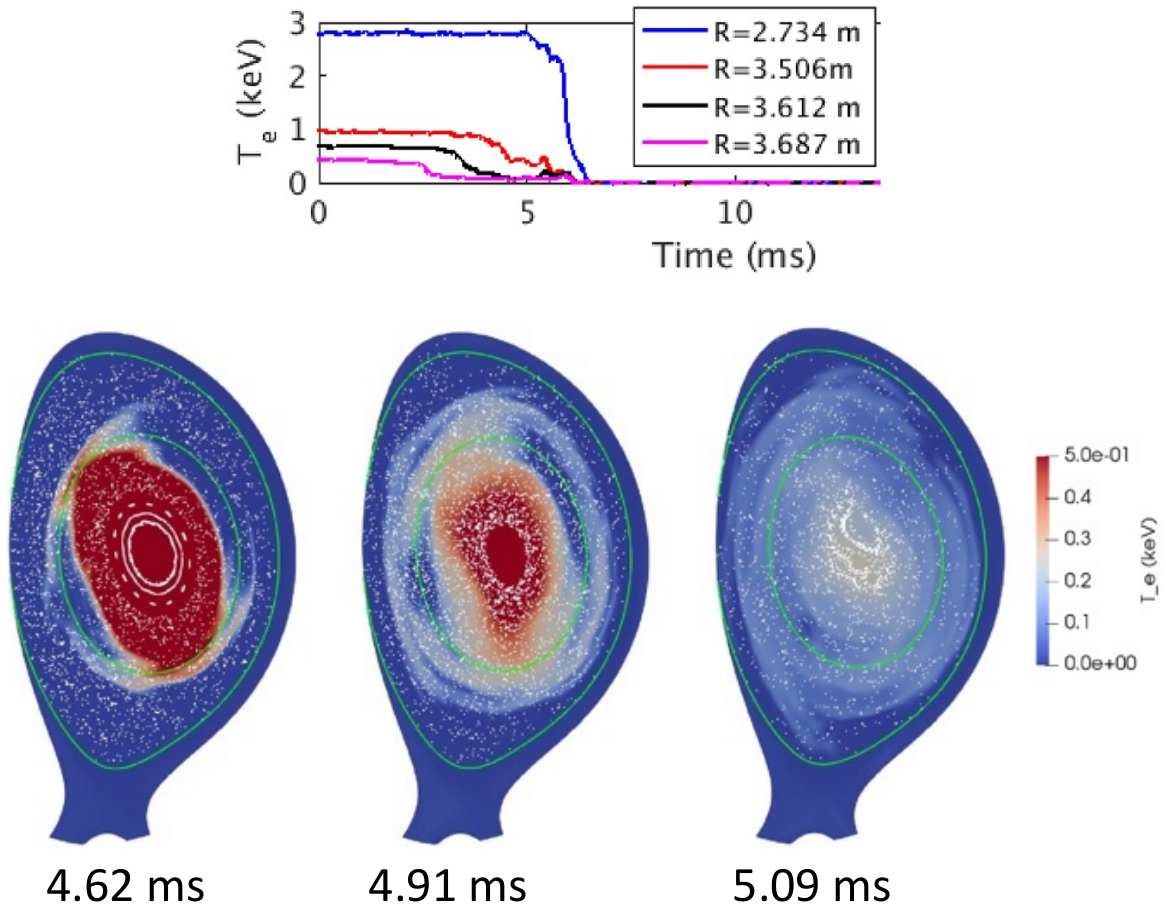} }
\caption{Large electron temperature differences persisted in a JET disruption initiated by a massive argon gas injection.  Time starts at the time of injection.  The two smooth green curves on the Poincar\'e plots are the pre-injection locations of the plasma edge and the $q=2$ surface.  This is a composite of two figures from E. Nardon, D. Hu, F. J. Artola, D Bonfiglio, et al, Plasma Phys. Control. Fusion \textbf{63}, 115006  (2021).  }  
\label{fig:T-diff}
\end{figure}

\color{black}

Valerie Izzo \cite{Izzo:2013} has carried out important simulations of the remarkably fast spreading of impurities after their injection in the DIII-D tokamak.  A simulation that is easier to compare with the $\vec{E}\times\vec{B}$ advective velocity and the diffusion that were calculated in Section \ref{sec:Plasma-ev} is the 2023 simulation of E. Nadon et al \cite{Nardon:2023}.  They reported a detailed analysis of a 2021 JOREK simulation \cite{Nardon:2021} of the formation of a current spike in a JET disruption that was triggered by   a massive gas injection of argon.  During the disruption, large electron temperature differences persisted even in regions of magnetic field line chaos, Figure \ref{fig:T-diff}.  \color{black}   The focus of their 2023 study was the flattening of $j_{||}/B$ along the chaotic magnetic field lines produced by the disruption.  They found consistency between their results and the flattening of $j_{||}/B$ by shear Alfv\'en waves that was proposed by Boozer \cite{Boozer:2019,Boozer:2020}.  But, the Nardon et al 2023 paper also found an $\vec{E}\times\vec{B}$ diffusive transport with a typical diffusion coefficient of 1000~m$^2$s$^{-1}$ and an $\vec{E}\times\vec{B}$ advective velocity of 5000~ms$^{-1}$, which were important for the spreading of the argon. The diffusion of 1000~m$^2$s$^{-1}$ agrees with the estimate of Equation (\ref{diffusion}) when $T_e=3~$keV and $B=3~$T, the approximate values in the pre-disruption plasma as given in \cite{Nardon:2021}.  The flow  given in Equation (\ref{flow}) would agree with the advective velocity reported in \cite{Nardon:2023}, 5000~ms$^{-1}$, if $a_T$ were 0.2~m.   

The consistency between the simulations of Nardon et al \cite{Nardon:2023} and the estimates in Section \ref{sec:Plasma-ev} of the effects of the electric field that enforces quasi-neutrality may seem paradoxical.  The actual equations being integrated in the simulation were given by Hu et al in \cite{Hu:JOREK2021} and do not include the relation between the electron pressure and the electric field.  However, Equation (7) of the Hu et al paper has the effect of enforcing $\vec{\nabla}\cdot\vec{j}=0$ in a multi-species plasma.  The uniqueness of the $\vec{E}_q$ in a Helmholtz decomposition implies that any valid method of imposing quasi-neutrality would yield the same result.

As discussed in \cite{Nardon:2023}, it is important that the physics associated with post-disruption chaotic magnetic fields be understood to address force, heat, and runaway electron loading on the walls.  Direct simulations are too challenging to allow parameter scans and have uncertainties that can only be addressed by a reliable physics understanding.


\section{Discussion \label{sec: Discussion}}

A deeper understanding of the results from simulations of tokamak disruptions, such as those of \cite{Izzo:2013,Nardon:2023,Nardon:2021,Jardin:fast breakup:2022,Sovenic:2019}, are of central importance to the success of ITER and the feasibility of tokamak power plants.  This understanding is also important for the analysis of non-resonant stellarator divertors and fast magnetic reconnection, not only in fusion but also in space and astrophysical plasmas.

The original divertor concept for the W7-X stellarator \cite{Strumbeger:divertor1992} used the magnetic field lines that strike the chamber walls but pass close to the outermost confining magnetic surface.  This is called a non-resonant divertor to distinguish it from the island divertor that was actually employed on W7-X \cite{W7-X_div2022}, which has the disadvantage of requiring a specific edge rotational transform.  The concepts of Cantori and turnstiles that are important for understanding magnetic surface breakup during tokamak disruptions are also an essential element in understanding non-resonant divertors \cite{Punjabi2022}.  Diffusion caused by the electric field required for quasi-neutrality also appears in the theory of non-resonant divertors \cite{Boozer:divetors2023}.   The observation of such effects in disruption simulations \cite{Nardon:2023} underscores their importance, but the primary code for studying stellarator divertors, EMC3-Eirene \cite{EMC3-Eirene} ignores electric fields.

Much of the theory and computation \cite{reconnection review} of magnetic reconnection is based on two-dimensional systems in which magnetic field line chaos is not mathematically possible.  The systems in which magnetic reconnection occurs in the laboratory, space, and astrophysics are fully three dimensional.  Many of the challenges to understanding magnetic reconnection are due to the historic focus on two-dimensional theory and are addressed by considering all three dimensions \cite{Boozer:B-ev}.   The magnetic reconnection studies that are a part of disruption simulations clarify the roles of chaos, Alfv\'en waves, quasi-neutrality potentials, and helicity conservation in the reconnection processes.

\section*{Acknowledgements}

The author thanks Eric Nardon for asking questions about earlier versions that lead to a much deeper understanding.

This material is based upon work supported by the U.S. Department of Energy, Office of Science, Office of Fusion Energy Sciences under Awards DE-FG02-03ER54696, DE-SC0018424, and DE-FG02-95ER54333. 

\section*{Author Declarations}

The author has no conflicts to disclose. \vspace{0.01in}


\section*{Data availability statement}

Data sharing is not applicable to this article as no new data were created or analyzed in this study.


\appendix 

\section{Magnetic helicity  \label{sec:helicity}} 

The concept of magnetic helicity was introduced in 1958 by Woltjer \cite{Woltjer}, but its importance as a conserved quantity even in highly turbulent plasmas was not appreciated until the 1974 paper of Bryan Taylor \cite{Taylor:1974}.   The equilibrium of a toroidal plasma requires magnetic helicity, $\int\vec{A}\cdot\vec{B}d^3x$, which can be produced by the external magnetic field, as in a stellarator, or by an internal plasma current, as in a tokamak.   The evolution of the helicity content of the region enclosed by the wall in a toroidal plasma  places a subtle constraint on the boundary condition at the wall in order for $\vec{E}_B$ to fully determine the magnetic field evolution.   

The form of the helicity evolution equation suggests it may be possible to maintain a tokamak in steady state by control of the electric potential on the vessel wall. The absence of a practical method of longterm control of the current makes tokamak steering to avoid disruptions seemingly impossible \cite{Boozer:steering}.  Unfortunately, a variation in the electric potential on the wall only allows current-density control on the magnetic field lines that strike the wall, which makes it of essentially no importance for disruption avoidance.

The three components of an arbitrary vector potential $\vec{A}(\vec{x},t)$ can be represented using a poloidal angle $\theta(\vec{x})$, a toroidal angle $\varphi(\vec{x})$, and a gauge function $g(\vec{x},t)$ as
\begin{equation}
\vec{A} = \psi\vec{\nabla}\frac{\theta}{2\pi} -  \psi_p(\psi,\theta,\varphi)\vec{\nabla}\frac{\varphi}{2\pi} +\vec{\nabla}g.  \label{A}\\
\end{equation}
As explained in the appendix to \cite{Boozer:RMP}, the only requirement is that $\vec{\nabla}\psi\cdot(\vec{\nabla}\theta \times \vec{\nabla}\varphi)= 2\pi \vec{B}\cdot\vec{\nabla}\varphi$ is neither zero nor infinity within the region of interest.

The helicity content \cite{Boozer:1986} $K_0$ in the region enclosed by the wall is
\begin{eqnarray}
K_0 &\equiv& \int \vec{A}\cdot\vec{B} d^3x \nonumber\\
&& - \left(\oint_w \vec{A}\cdot \frac{\partial \vec{x}}{\partial\theta}\frac{d\theta d\varphi}{2\pi}\right) \left(\oint_w \vec{A}\cdot \frac{\partial \vec{x}}{\partial\varphi}\frac{d\theta d\varphi}{2\pi}\right) \nonumber\\
&=&2\int_0^\Psi d\psi \oint (\Psi_p - \psi_p) \frac{ d\theta d\varphi}{(2\pi)^2}, \mbox{  where   }  \label{K_0}\\
\Psi &\equiv& \oint_w \vec{A}\cdot \frac{\partial \vec{x}}{\partial\theta}\frac{d\theta d\varphi}{2\pi} =\int \frac{\vec{B}\cdot\vec{\nabla}\varphi }{2\pi}d^3x, \mbox{  and  }\\
\Psi_p &\equiv& \oint_w \vec{A}\cdot \frac{\partial \vec{x}}{\partial\varphi}\frac{d\theta d\varphi}{2\pi}  
\end{eqnarray}  
The poloidal and the toroidal loop integrals of the vector potential  $\vec{A}$ in Equation (\ref{K_0}) remove the indeterminacy of the gauge, $g$, from the definition of the helicity.   Reference \cite{Boozer:1986} omitted the requirement for an integration over both $\theta$ and $\varphi$ in these integrals. $\Psi$ is called the total toroidal flux, and $\Psi_p$ is called the total poloidal flux enclosed by the wall.

Following \cite{Boozer:1986}, the time derivative of the helicity content is
\begin{eqnarray}
\frac{dK_0}{dt} &= & 2V_s\Psi - 2 \int \vec{E}\cdot\vec{B}d^3x - 2 \oint_w \Phi_w\vec{B}\cdot d\vec{a}. \hspace{0.3in} \label{dK_0/dt}
\end{eqnarray}
 $V_s$ is the toroidal loop voltage produced by a solenoid located in the hole of the torus.  Mitchell Berger wrote a well known review of the helicity and its time dependence \cite{Berger}, but he and other authors omit the loop voltage from the central solenoid, $V_s$.

The general expression for an electric field, Equation (\ref{E eq}) can be substituted in Equation (\ref{dK_0/dt}) for the time derivative of the helicity content.  The result is 
\begin{eqnarray}
\frac{dK_0}{dt}&=& 2V_s\Psi - 2\int \frac{V_\ell}{2\pi} \vec{B}\cdot\vec{\nabla}\varphi d^3x \\
&=& 2 \int_0^\Psi d\psi \oint (V_s-V_\ell) \frac{d \theta d\varphi}{(2\pi)^2}. \label{k_0-simple}
\end{eqnarray}

The subtlety in Equation (\ref{k_0-simple}) is the loop voltage.  The potential $\Phi$ obeys the differential equation
\begin{eqnarray}
\frac{\partial\Phi}{\partial \varphi} = - \frac{\vec{E}\cdot\vec{B}}{\vec{B}\cdot\vec{\nabla}\varphi} + \frac{V_\ell}{2\pi}
\end{eqnarray}
with $V_{\ell}$ chosen to make $\Phi$ a single-valued function of $\varphi$.  The magnetic field lines are of two types.  The first type is confined within the region enclosed by the wall and never strike it.  For these lines, the loop voltage is defined by the integral
\begin{equation}
V^{(c)}_\ell =  \lim_{N_t\rightarrow\infty} \int_0^{2\pi N_t} \frac{\vec{E}\cdot\vec{B}}{\vec{B}\cdot\vec{\nabla}\varphi} \frac{d\varphi}{N_t},
\end{equation}
where $N_t$ is the number of toroidal circuits included in the integration.

The second type of magnetic field line strikes the wall in two locations and are called open lines.  The line enters the plasma region at a place where $\vec{B}\cdot\hat{n}<0$ but must after a sufficiently large number of toroidal circuits, $N_t$, strike the wall in a location where $\vec{B}\cdot\hat{n}>0$.  $N_t$ can be arbitrarily large, but if it were infinity the line would have to have penetrated a magnetic surface, which is impossible.   A magnetic surface is any surface on which a field line can be followed forwards or backwards forever and never leave. A field line can enter an almost isolated chaotic region and remain for an arbitrarily large number of toroidal circuits, but $\vec{\nabla}\cdot\vec{B}=0$ implies that if there is a way in there must be a way out. 

When $\Phi_w^{in}$ is the potential on the wall at the location where the line came into the plasma region and $\Phi_w^{out}$ is the potential on the wall at the location where the line went out of the plasma region, 
\begin{equation}
V^{(o)}_\ell =  \Phi_w^{out} - \Phi_w^{in} + \int_{in}^{out} \frac{\vec{E}\cdot\vec{B}}{\vec{B}\cdot\vec{\nabla}\varphi} d\varphi. \label{wall loop}
\end{equation}
The boundary condition $\Phi_q]_w=0$ eliminates $\Phi_q$ from Equation (\ref{wall loop}) and ensures $\Phi_q$ has no direct effect on the evolution of the magnetic field.

When the plasma obeys a simple Ohm's law, $\vec{E}\cdot\vec{B}=\eta_{||} \vec{j}\cdot\vec{B}$, and the loop voltage for bound lines is dissipative.  However, if the difference in the potential on the wall between places where $\vec{B}\cdot\hat{n}>0$ and places where $\vec{B}\cdot\hat{n}<0$ is sufficiently large the loop voltage $V^{(o)}_\ell$ can increase the magnitude of the helicity.

The average loop voltage due to the wall potential $\Phi_w$ in the region that carries a toroidal flux $\Psi_o$ in open lines is 
\begin{eqnarray}
V_w \equiv \frac{\oint \Phi_w \vec{B}\cdot d\vec{a}}{\Psi_o}.
\end{eqnarray}
The contribution the wall potential to helicity evolution is $-2\int V_w d\psi =- 2\oint \Phi_w \vec{B}\cdot d\vec{a}$, which equals the last term in Equation (\ref{dK_0/dt}).   An expression for helicity evolution close to the conventional one is obtained with the plasma loop voltage $V_\ell^{(p)}$ defined as
\begin{eqnarray}
V_\ell^{(p)}& \equiv& V^{(c)}_\ell  \mbox{   when a field line is confined }\\
&\equiv& \int_{in}^{out} \frac{\vec{E}\cdot\vec{B}}{\vec{B}\cdot\vec{\nabla}\varphi} d\varphi \mbox{ when it is not. } \\
\frac{dK_0}{dt} &=& 2  \oint (V_s-V_\ell^{(p)}) \frac{d\psi d \theta d\varphi}{(2\pi)^2} \nonumber \\
&& \hspace{0.8in} - 2\oint \Phi_w \vec{B}\cdot d\vec{a}. \hspace{0.2in}
\end{eqnarray}

The spatial and temporal variation in the electric potential $\Phi_w$ on the wall could be  controlled using the $\eta \vec{j}$ electric fields that would result from relatively small externally-driven currents in the wall.  Although this could maintain the helicity in a tokamak,  the resulting situation would not be desirable.  The spatial variation in $\Phi_w$ would only drive currents on the open field lines, and the plasma current in the region confined by the outermost magnetic surface would decay until either all of the magnetic surfaces disappeared or a helicity-conserving disruption opened all of the magnetic field lines.  In either case $j_{||}/B$ would relax to a constant throughout the chamber, which would remove the drive for kinks.  If $\vec{B}\cdot\hat{n}$ on the wall were essentially axisymmetric, magnetic surfaces would naturally re-form with a higher plasma temperature in the core, which would give slower helicity decay there.  By careful control of the time dependence of $\Phi_w$, the helicity could be held fixed but with a periodic breaking and re-formation of magnetic surfaces.

A method of controlling the current profile seems a requirement for tokamak power plants.  The suggestion of Jensen and Chu \cite{Jensen:1984} that this could be done by a the variation in $\Phi_w$ meant that this effect should be carefully studied.  Unfortunately, we have shown that although the magnetic helicity could be held to a steady-state value that the resulting tokamak is time varying and very undesirable.



\end{document}